%% file: RandI_paper.tex
\documentclass[journal,10pt,onecolumn]{IEEEtran}
\usepackage{amssymb,amsmath,graphicx}
\usepackage{amsmath,mathrsfs,upgreek}
\usepackage{algorithmicx}
\usepackage{algpseudocode}
\usepackage{mdframed}
\usepackage{xspace}
\usepackage{graphicx}
\usepackage{graphics}
\usepackage{caption}
\usepackage{subcaption}
\usepackage{subfig}
\usepackage{multirow}
\usepackage{epstopdf}
\usepackage{array}
\usepackage{cite}
\usepackage{color}
\usepackage{float}
\usepackage{tikz}
\usepackage{tikz-cd}
\usetikzlibrary{calc,fit,arrows}
\usetikzlibrary{positioning}
\usetikzlibrary{shapes}
\renewcommand\eqref[1]{(\ref{#1})}
\usepackage{fancyhdr}
\fancypagestyle{customplain}{%
\fancyhf{}%
\fancyhf[cf]%
}

\makeatletter
\let\old@ps@headings\ps@headings
\let\old@ps@IEEEtitlepagestyle\ps@IEEEtitlepagestyle
\def\confheader#1{%
  \def\ps@headings{%
    \old@ps@headings%
    \def\@oddhead{\strut\hfill#1\hfill\strut}%
    \def\@evenhead{\strut\hfill#1\hfill\strut}%
  }%
  \def\ps@IEEEtitlepagestyle{%
    \old@ps@IEEEtitlepagestyle%
    \def\@oddhead{\strut\hfill#1\hfill\strut}%
    \def\@evenhead{\strut\hfill#1\hfill\strut}%
  }%
  \ps@headings%
}
\makeatother

\confheader{%
  Submitted to 19th International Conference on Information Fusion, Germany, July 2016
}

\ifCLASSINFOpdf
\else
\fi
\IEEEoverridecommandlockouts

\begin{document}
%
\title{\huge Object Recognition and Identification Using ESM Data}
\author{\IEEEauthorblockN{\small  E. Taghavi, D. Song, R. Tharmarasa, T. Kirubarajan\\}
\IEEEauthorblockA{\small Department of Electrical and Computer Engineering\\
McMaster University, Hamilton, Ontario, Canada\\
$\{${taghave,songd8,tharman,kiruba}$\}$@mcmaster.ca\\}
\and
\IEEEauthorblockN{\small Anne-Claire Boury-Brisset$^{*}$\thanks{$^{*}$Anne-Claire Boury-Brisset is with DRDC Valcartier, Quebec, Canada.}, Bhashyam Balaji$^{\dagger}$\thanks{$^{\dagger}$Bhashyam Balaji is with DRDC Ottawa, Ontario, Canada.}\\}
\IEEEauthorblockA{\small Defence Research and Development Canada\\
$\{${anne-claire.boury-brisset,bhashyam.balaji}$\}$@drdc-rddc.gc.ca}
}
%

%


\maketitle
\pagestyle{plain}
\begin{abstract}
Recognition and identification of unknown targets is a crucial task in surveillance and security systems. Electronic Support Measures (ESM) are one of the most effective sensors for identification, especially for maritime and air--to--ground applications. In typical surveillance systems multiple ESM sensors are usually deployed along with kinematic sensors like radar. Different ESM sensors may produce different types of reports ready to be sent to the fusion center. The focus of this paper is to develop a new architecture for target recognition and identification when non--homogeneous ESM and possibly kinematic reports are received at the fusion center. The new fusion architecture is evaluated using simulations to show the benefit of utilizing different ESM reports such as attributes and signal level ESM data. 
\end{abstract}


%
\IEEEpeerreviewmaketitle

\input{introduction.tex}

\input{frameworks.tex}

\input{problem.tex}

\input{systems.tex}

\input{discussion.tex}

\input{conclusions.tex}

\bibliographystyle{plain}
\bibliography{references}

\end{document}

%% file: introduction.tex
\section{Introduction}
Processing Electronic Support Measure (ESM) data in surveillance systems \cite{gamma2001ew101,poisel2004target,cheng2012electronic} is one of the fundamental tasks in Recognition and Identification (R\&I) \cite{matuszewski2014knowledge,kawalec2004specific}. Identifying the emitter type based on the signals received from an ESM sensor has been addressed in the literature \cite{davies1982automatic,roe1990knowledge,kawalec2004specific}. The approaches in these previous works are suitable for systems with a single ESM sensor. In modern surveillance systems, the combination of radar reports and ESM is used to identify the targets and to further classify them into known platforms \cite{challa2001joint,demers2006identity,li2007esm}. These systems benefit from the features that can be extracted from the kinematic \cite{bar2001estimation} of the targets as well as those of from ESM to classify the targets with higher accuracy. With the advances in recent years, the number of sensors has multiplied in modern surveillance systems and the need for new sensor fusion architectures has been identified. In \cite{gad2002data}, a comprehensive categorization of possible levels of processing that can be utilized in different applications such as air--to--ground and ground--to--ground surveillance was given. 

The primary focus of this paper is to handle multiple ESM sensors along with radar (kinematic) measurements to improve R\&I accuracy. Previous works have addressed design of algorithms for processing and fusing ESM data in different contexts. In \cite{smestad2001esm}, ESM sensors and their capabilities are investigated for air defence system design. The concept of attribute estimation is discussed for multisensor scenarios in \cite{bar2011tracking}. Going beyond the traditional architectures for R\&I systems, in this paper, a new approach for processing ESM and kinematic data is proposed to achieve the goals introduced in standards such as STANAG $4162$ \cite{stanag20014162}. Although the Bayesian framework \cite{bernardo2001bayesian} is mainly considered in most R\&I systems, this paper considers other frameworks such as evidence theory \cite{zadeh1986simple} and possibility theory and fuzzy sets \cite{zadeh1978fuzzy}, whenever possible, in addition to probabilistic solutions in different blocks of the new hierarchy, to achieve a better utilization of available information.

If multiple ESM sensors are available in a surveillance system, there may be a variety of information in their output. For example the ESM data sent to the fusion center may range from raw (unprocessed) ESM data to fully processed information such as threat category or lethality \cite{hall1997introduction} of a specific target. Some of these data can be considered as probabilistic distributions while others may be in the form of sets and intervals possibly with probabilities assigned to them. The fusion of heterogeneous data for R\&I and the management of the flow of information in the scenarios is the focus of this paper. This development motivates the new architecture for fusing all information to better identify the targets.

In order to understand the problem, the traditional frameworks that has been used to process ESM data are reviewed in Section \ref{frameworks}. In Section \ref{problem}, the fusion of data from a single ESM and radar sensor is considered. Section \ref{new_idea} is devoted to the new system design and  to show how it can be used in modern ESM fusion systems. In Section \ref{discussion}, a simple example is considered with discussions on possible issues in the design of the proposed architecture. Finally, conclusions and future works are discussed in Section \ref{conclusion}.

%% file: frameworks.tex
\section{Review of frameworks for R\&I} \label{frameworks}
There are different R\&I frameworks one can use in surveillance systems. In advanced systems, the ability to take advantage of different frameworks is advantageous. Due to this emergence of multiple fusion approaches and in order to further discuss and formulate the ideas that are necessary for the development of the new architecture proposed in this paper, a brief review of common fusion frameworks for R\&I is given in this section. 

\subsection{Probability theory -- Bayesian framework}
Bayesian framework refers to the classical method that uses the Bayes' theorem in statistical inference, which includes testing hypotheses and deriving estimates. The core of the Bayesian framework is to update the probability of a hypothesis or the probability distribution of a random variable when evidence (measurement) becomes available based on the Bayes' formula \cite{bar2001estimation}:
\begin{IEEEeqnarray}{rCl}
	p(B_i|y) &=& \frac{p(y|B_i)p(B_i)}{p(y)}	\nonumber \\
	&=& \frac{p(y|B_i)p(B_i)}{\sum_{i=1}^{n}{p(y|B_i)p(B_i)}}
\end{IEEEeqnarray}
for hypothesis (event) and
\begin{IEEEeqnarray}{rCl}
	p(x|y) &=& \frac{p(y|x)p(x)}{p(y)}	\nonumber\\
	&=& \frac{p(y|x)p(x)}{\int{p(y|x)p(x)dx}}
\end{IEEEeqnarray}
for random variables.
In the above, $P(B_i|y)$ and $p(x|y)$ are the posterior probabilities and, $p(B_i)$ and $p(x)$ are the prior probabilities. Bayesian framework is particularly useful when evidence is received sequentially. In this case, the posterior probability becomes the next prior probability at the next time instant.

\subsection{Evidence theory -- Belief functions}
In probability distributions, beliefs are distributed over the elements of the outcome space. In contrast, for a belief function, the distribution, also known as mass function, attributes belief to subsets of the outcome space. In other words, if there is belief attributed to a subset, there is a reason to believe that the outcome will be among the elements of that subset, without committing to any preference among those elements. One can define a mass distribution $m_\Theta$ as a mapping from subsets of a frame of discernment $\Theta$ into the unit interval:
\begin{eqnarray}
m_{\Theta} & : & 2^{\Theta}\mapsto\left[0,1\right]\label{eq:ehsan_mtheta}
\end{eqnarray}
such that
\begin{eqnarray}
m_{\Theta}\left(\emptyset\right)=0 & \mathrm{\; and\;} & \sum_{A\subseteq\Theta}m_{\Theta}\left(A\right)=1
\end{eqnarray}
In this context, focal elements are defined as subsets that are attributed to non--zero mass. It is worth noting that the belief in a hypothesis $A\left(A\subseteq\Theta\right)$ is constrained to lie within an interval $\left[\mathrm{Bel}\left(A\right),\mathrm{Pl}(A)\right]$, where
\begin{eqnarray}
\mathrm{Bel}\left(A\right)=\sum_{A_{i}\subseteq A}m_{\Theta}\left(A_{i}\right) & ; & \mathrm{Pl}(A)=1-\mathrm{Bel}\left(\urcorner A\right)\label{eq:ehsan_BelPl}
\end{eqnarray}
These bounds are known as belief (support) and plausibility, respectively.
To combine two different belief functions together, one can use the Dempster's rule of combination \cite{yager1987dempster} as the fusion operator. Note that this rule finds the common shared belief between different sources and ignores all the conflicting belief through a normalization factor. Therefore, the combination (joint mass) of two sets of masses $m_1$ and $m_2$ can be written as
\begin{eqnarray}
m_{1,2}\left(\ensuremath{\emptyset}\right) & = & 0\label{eq:ehsan_m12}\\
m_{1,2}\left(A\right) & = & \left(m_{1}\oplus m_{2}\right)\left(A\right)\nonumber \\
 & = & \frac{1}{1-K}\sum_{B\cap C=A\neq\emptyset}m_{1}\left(B\right)m_{2}\left(C\right)\label{eq:ehsan_m12_full}
\end{eqnarray}
where $K$ is the measure of the amount of conflict between the two mass sets and is defined by
\begin{eqnarray}
K & = & \sum_{B\cap C=\emptyset}m_{1}\left(B\right)m_{2}\left(C\right)\label{eq:ehsan_K}
\end{eqnarray}

\subsection{Possibility and fuzzy set theory}
To further process the information that has linguistic characteristics, the idea of fuzzy sets and possibility theory can be used \cite{zadeh1978fuzzy}. If possibility distributions are defined as fuzzy restrictions, then the imprecision in natural languages, which is possibilistic rather than probabilistic, can be addressed in a systematic manner. This subject is beyond the scope of this paper and will be explained in the work in progress for a general framework of ESM data fusion.

%% file: problem.tex
\section{Problem Formulation} \label{problem}
\begin{figure*}[htbp!]
\centering
\tikzstyle{int}=[draw, fill=blue!20, rounded corners,minimum size=2em]
\tikzstyle{init} = [pin edge={to-,thin,black}]
\tikzstyle{line} = [draw, -latex']
\tikzstyle{decision} = [diamond, draw, fill=blue!20, 
    text width=2cm, text badly centered, node distance=3cm]

\begin{tikzpicture}[node distance=2.5cm,auto,>=latex']
    \node [int] (esm) {ESM};
    \node [int] (attr1) [right of=esm , node distance=2.5cm] {Attributes};
    \node (recog) [int , pin={[init]above:{ESM Features Set}} , right of=attr1 , node distance=3cm] {Recognition};
    
    \node [int] (meas) [below of=esm , text width = 2cm , text centered] {{Kinematic Measurement}};
    \node [int] (tracker) [right of=meas , node distance=3cm] {Tracker};
    \node (recog2) [int , pin={[init]below:{Tracker Features Set}} , right of=tracker, node distance=3cm] {Recognition};

    \node [int] (assoc) [below right= 0.5cm and 1cm of recog , text width = 3cm , text centered] {Classifier};
    \node [coordinate] (end) [right of=assoc, node distance=3cm]{};

    \path[->] (esm) edge node { $\{\mathbf{y},\mathbf{U}\}$ } (attr1);
    \path[->] (attr1) edge node { $\{\mathbf{a},\mathbf{A}\}$ } (recog) ;

    \path[->] (meas) edge node { $\{\mathbf{z},\mathbf{R}\}$ } (tracker);
    \path[->] (tracker) edge node[sloped, anchor=center, below] {$\{\mathbf{\mu},\mathbf{\Lambda}\}$}  (recog2) ;
    \path[->] (tracker) edge node[sloped, anchor=center, above] {$\{\mathbf{x},\mathbf{P}\}$}  (recog2) ;
  
    \path [line] (recog) -| node[xshift=-2cm,yshift=0.2cm] {$\{\mathbf{r}_{esm}$,$\mathbf{\rho}_{esm}\}$} (assoc);
    \path [line] (recog2) -| node[yshift=0.2cm] {$\{\mathbf{r}_{pos}$,$\mathbf{\rho}_{pos}\}$} (assoc);
    \path[->] (assoc) edge node { $\{\mathbf{c}$,$\mathbf{C}\}$ } (end) ;

    \node[fit= (tracker) (recog2), dashed,draw,inner sep=0.15cm,rounded corners] (Box) {};

\end{tikzpicture}
\caption{Block diagram of ESM/Kinematic fusion}
\label{fig_block}
\end{figure*}
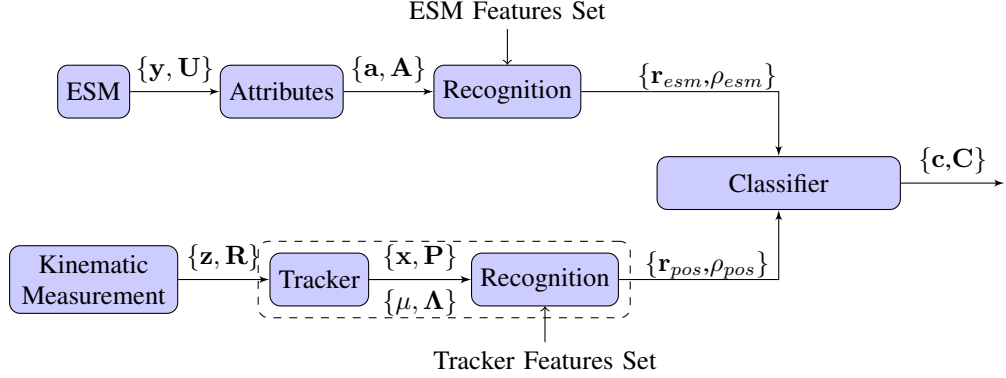

Classification of targets into different categories is one of the most important tasks in Recognition and Identification. There are works in literature on different ways for the classification of data in distributed tracking systems for the purpose of recognition and identification \cite{bilik2006gmm,brooks2003distributed}. In this paper we focus on addressing the issue of the fusion of ESM and kinematic measurements at three different levels. Figure \ref{fig_block} shows a general block digram on how the classification process is carried out based on ESM and kinematic measurements.

For air--to--ground target tracking, if altimeter  or terrain data is available, an estimate of the ground or sea level targets can be found by using a single ESM sensor alone \cite{collins1989terrain}. This estimate has the advantage of being uncorrelated with data processed for estimating track features and further can be fused with kinematic estimate of other sensors.

In Figure \ref{fig_block}, output types can be numerical, objective, structures, low--level (data), etc., assuming that they are hard data. Due to the existence of various types of data, such data should be defined within the proposed framework so that each block can use the data it receives. If soft data such as operator reports are available, qualitative, subjective, unstructured and higher-level (information) data also must be included within the framework. Imperfect nature of information can also be considered in the architecture. These imperfections in data can come from different sources such as the uncertainty in the information, imprecision and the conflict between the processed data.  

\subsection{ESM measurement}
Electronic support measures consist of various measurements and depend on the sensors that are mounted on the platform. To give a better view of the problem, in this paper, it is assumed that ESM measurements are Pulse Repetition Interval (PRI) high, PRI low, frequency high, frequency low, Pulse Width (PW) high, PW low and amplitude. Table \ref{tab_ESM} gives a short description of these measurements.
\begin{table}[htbp!]
\caption{ESM measurement space}
\label{tab_ESM}
\begin{centering}
\centering{}%
\begin{tabular}{|c|c|}
\hline 
ESM mseasuement & Mean\tabularnewline
\hline 
\hline 
PRI High & $\kappa_{h}$\tabularnewline
\hline 
PRI Low & $\kappa_{l}$\tabularnewline
\hline 
Frequency High & $f_{h}$\tabularnewline
\hline 
Frequency Low & $f_{l}$\tabularnewline
\hline 
PW High & $\delta_{h}$\tabularnewline
\hline 
PW Low & $\delta_{l}$\tabularnewline
\hline 
Amplitude & $\alpha$\tabularnewline
\hline 
\end{tabular}
\par\end{centering}
\end{table}
All these measurements are perturbed with white noise of known distributions with means defined in Table \ref{tab_ESM}. Measured sensor data can be sent to the next block for further processing via a vector and its associated distribution model parameters as $\left\{ \mathbf{y},\mathbf{U}\right\}$.

\subsection{Attribute estimate}
Target attributes are various variables with discrete values that primarily give information to distinguish between the targets by characterizing the measurements into different groups by assigning probabilities to possible outcomes \cite{bar2011tracking}.
Assuming $\left\{ \mathbf{y},\mathbf{U}\right\}$ is received, the measurement can be characterized by the conditional probability density function (pdf) for $\left\{ a_{1},\ldots,a_{n}\right\} $ attributes as
\begin{eqnarray}
p\left(y\mid a=a_{i}\right) & = & g_{i}\left(y\right)
\label{eq:ehsan_att}
\end{eqnarray}
where $g_{i}$ is the likelihood function of the attribute $a_i$. In addition, prior information of the attributes can be modeled as
\begin{eqnarray}
p_{i}\left(0\right) & = & \mathbb{P}\left\{ a=a_{i}\right\} \hspace{1cm}i=1,\ldots,n\label{eq:ehsan_prior}
\end{eqnarray}
Starting with processing the first measurement, the conditional probability of the target having attribute $i$ is, in the Bayesian framework
\begin{eqnarray}
p_{i}\left(1\right) & \overset{\bigtriangleup}{=} & \mathbb{P}\left\{ a=a_{i}\mid y(1)\right\} =\frac{1}{c_{1}}g_{i}\left(y(1)\right)p_{i}\left(0\right)\label{eq:ehsan_p1}
\end{eqnarray}
where the normalization factor $c_1$ is
\begin{eqnarray}
c_{1} & = & \sum_{j=1}^{n}g_{j}\left(y(1)\right)p_{j}\left(0\right)\label{eq:ehsan_c1}
\end{eqnarray}
If more measurements are available through different time steps, the recursive updatimg of attribute $i$ can be written as
\begin{eqnarray}
p_{i}\left(k\right) & = & \frac{1}{c_{k}}g_{i}\left(y(k)\right)p_{i}\left(k-1\right)\label{eq:ehsan_pk}
\end{eqnarray}
where 
\begin{eqnarray}
c_{k} & \overset{\bigtriangleup}{=} & \sum_{j=1}^{n}g_{j}\left(y(k)\right)p_{j}\left(k-1\right)\label{eq:ehsan_ck}
\end{eqnarray}
Note that it is assumed the measurements are independent conditioned on the attribute. Then, the so called attribute estimate can be shown as a list of probabilities assigned to different attributes. Based on practical surveillance systems, informative attributes with possible outputs are shown in Table \ref{tab_att}. Each probability or evidence data presented with $\Xi{(\cdot)}$ can be a list of scalars or intervals with their associated probabilities or probability density functions. The choice of the data type depends on the application and model of the sensor.
\begin{table}
\caption{ESM attributes}
\label{tab_att}
\begin{centering}
\begin{tabular}{|c|c|}
\hline 
Attribute & $\mathbb{P}\left\{ \mathcal{S}\right\} $\tabularnewline
\hline 
\hline 
Speed & $\Xi_{\mathrm{Speed}}$\tabularnewline
\hline 
Shape & $\Xi_{\mathrm{Shape}}$\tabularnewline
\hline 
Length & $\Xi_{\mathrm{Length}}$\tabularnewline
\hline 
Size & $\Xi_{\mathrm{Size}}$\tabularnewline
\hline 
Emitter ID & $\Xi_{\mathrm{ID}}$\tabularnewline
\hline 
Number of emitters & $\Xi_{\mathrm{Nb}}$\tabularnewline
\hline 
\end{tabular}
\par\end{centering}
\end{table}

\subsection{Recognition of attributes estimates}
In order to further process the data received at this stage, i.e., $\{\mathbf{a},\mathbf{A}\}$, it is necessary to use a framework that can combine different types of information in $\{\mathbf{a},\mathbf{A}\}$. For example, to find out the threat category of a target, which can be friendly or foe, one should combine the information of the shape and speed together. If this information is in the form of lists, one can use belief functions and combination rules to fuse and later classify the data \cite{zadeh1986simple,yager1987dempster,ristic2011target,ristic2005target,ristic2005targetA}. If all the data is in terms of probabilistic models, Bayesian framework can be adopted for the purpose of classification \cite{williams1998bayesian}. In the case of a combination of belief and probabilistic functions, approximations can be applied to belief functions as stated in \cite{voorbraak1989computationally} to be compatible with the Bayesian framework. In all cases, it is important to design the model such that uncertainty information is also generated in addition to recognition and classification results.

\subsection{Position measurement}
In tracking applications, the target position is mainly measured by a radar or sonar, in polar or spherical coordinates \cite{bar2001estimation,Aidala1979}. However, target motion is usually best modeled in Cartesian coordinates. To cope with different coordinates used in measurement and target motion space, the received measurement is converted into Cartesian coordinates prior to tracking by the method proposed in \cite{Bordonaro2014}. Assume $\mathbf{z}_m=[r_m\;\; \theta_m]'$ is the measurement in the two-dimensional case where $r_m$ and $\theta_m$ are the range and bearing measurement perturbed by independent Gaussian noise with a zero mean and standard deviation $\sigma_r$ and $\sigma_\theta$, respectively. Then, the converted measurement in Cartesian coordinates can be expressed as
\begin{IEEEeqnarray}{rClCl}
    \mathbf{z} &\triangleq& \begin{bmatrix} x_m\\y_m \end{bmatrix} &=& e^{\sigma^2_\theta/2} \begin{bmatrix} r_m\cos{\theta_m}\\r_m\sin{\theta_m} \end{bmatrix}
\end{IEEEeqnarray}
and its associated covariance matrix is
\begin{IEEEeqnarray}{rCl}
    \mathbf{R} &\triangleq& \begin{bmatrix} R_{11} & R_{12} \smallskip\\ R_{21} & R_{22} \end{bmatrix} \label{cov_z}
\end{IEEEeqnarray}
where 
\begin{IEEEeqnarray}{rCl}
    R_{11} &=& \frac{1}{2}(r_m^2+\sigma_r^2)\left[ 1+\cos{(2\theta_m)}e^{-2\sigma_\theta^2} \right] \nonumber \\
    &&+\left[ e^{\sigma_\theta^2}-2 \right]r_m^2\cos^2{\theta_m} \\
    R_{22} &=& \frac{1}{2}(r_m^2+\sigma_r^2)\left[ 1-\cos{(2\theta_m)}e^{-2\sigma_\theta^2} \right] \nonumber \\
    &&+\left[ e^{\sigma_\theta^2}-2 \right]r_m^2\sin^2{\theta_m} \\
    R_{12} &=& \frac{1}{2}(r_m^2+\sigma_r^2)\left[ \sin{(2\theta_m)}e^{-2\sigma_\theta^2} \right] \nonumber \\
    &&+\left[ e^{\sigma_\theta^2}-2 \right]r_m^2\sin{\theta_m}\cos{\theta_m} \\
    R_{21} &=& R_{12}
\end{IEEEeqnarray}

\subsection{Tracker estimate}
Assume that each target belongs to one of $s$ known classes $c\in\{1,2,\ldots,s\}$ and the kinematic behavior of each target class is characterized by a linear dynamical model set $\mathcal{S}_c$ with Markov property and corresponding transition matrix $\mathcal{P}_c$. The $i$th dynamical model $M^i_c \in \mathcal{S}_c, i=1,...,r(c)$ can then be represented by
\begin{IEEEeqnarray}{rCl}
    \mathbf{x}(k) &=& F(M^i_c)\mathbf{x}(k-1)+G(M^i_c)\mathbf{u}(k,M^i_c)+\mathbf{v}(k,M^i_c) \IEEEeqnarraynumspace
\end{IEEEeqnarray}
where $\mathbf{x}(k)\triangleq[x(k)\;\,\dot{x}(k)\;\,\ddot{x}(k)\;\,y(k)\;\,\dot{y}(k)\;\,\ddot{y}(k)]'$ is target kinematic state at time $k$ and $\mathbf{v}(k,M^i_c)$ is a white independent identically distributed (iid) Gaussian process noise with zero mean and covariance $Q(m^i_c)$. Then the measurement equation can be written as
\begin{IEEEeqnarray}{rCl}
    \mathbf{z}(k) &=& H\mathbf{x}(k)+\mathbf{w}(k)
\end{IEEEeqnarray}
where $\mathbf{w}(k)$ is assumed as iid Gaussian noise with zero mean and covariance $\mathbf{R}$ given in \eqref{cov_z}.

Here it is assumed that there are $s$ number of IMM estimators \cite{bar2001estimation} corresponding to $s$ target classes, respectively. These $s$ number of IMM estimators are set up for each target. The state estimate $\hat{\mathbf{x}}_i(k|k)$, associated covariance matrix $P_i(k|k)$, posterior probability of mode $j$ being correct given the measurement up to time $k-1$, noted as $\mu_i^j(k)$, and the measurement likelihood function conditioned on mode $j$, noted as $\Lambda_i^j(k)$, obtained by each IMM estimator $i=1,...,s$ and are provided to the subsequent recognition and fusion steps. 

\subsection{Recognition of tracker output}
Given the measurement sequence $\mathbf{Z}^k = \{\mathbf{z}(0),\ldots,\mathbf{z}(k)\}$ up to time $k$, the posterior probability of target being class $i$, noted as $P(c=i|\mathbf{Z}^k)$, in the recursive Bayesian framework, is given by
\begin{IEEEeqnarray}{rCl}
    P(c=i|\mathbf{Z}^k) &=& P(c=i|\mathbf{z}(k),\mathbf{Z}^{k-1}) \nonumber \smallskip\\
    &=& \frac{P(\mathbf{z}(k)|c=i,\mathbf{Z}^{k-1})P(c=i|\mathbf{Z}^{k-1})}{\sum_{j=1}^{s}{P(\mathbf{z}(k)|c=j,\mathbf{Z}^{k-1})P(c=j|\mathbf{Z}^{k-1})}}   \nonumber \\*
\end{IEEEeqnarray}
where 
\begin{IEEEeqnarray}{rCl}
    P(\mathbf{z}(k)|c=i,\mathbf{Z}^{k-1}) &=& \sum_{j=1}^{r(i)}P(\mathbf{z}(k)|u(k)=M_i^j,c=i,\mathbf{Z}^{k-1})     \nonumber \\
    && \times P(u(k)=M_i^j|c=i,\mathbf{Z}^{k-1})    \nonumber \\
    &=& \sum_{j=1}^{r(i)}\mu_i^j(k)\Lambda_i^j(k)
\end{IEEEeqnarray}
Also, $\mu_i^j(k)$ is the posterior probability of mode $j$ being correct given the measurement up to time $k-1$ and $\Lambda_i^j(k)$ is the likelihood function conditioned on mode $j$ of class $i$ at time $k$. Finally, we have
\begin{IEEEeqnarray}{rCl}
    P(c=i|\mathbf{Z}^k) &=& \frac{\sum_{j=1}^{r(i)}\mu_i^j(k)\Lambda_i^j(k)P(c=i|\mathbf{Z}^{k-1})}{\sum_{j=1}^{s}\left\{ \sum_{m=1}^{r(j)}\mu_j^m(k)\Lambda_j^m(k)P(c=j|\mathbf{Z}^{k-1})\right\}} \nonumber \\*
\end{IEEEeqnarray}
and
\begin{IEEEeqnarray}{rCl}
    P(c=i|\mathbf{Z}^0) &=& P_0(c=i)
\end{IEEEeqnarray}
where $P_0(c=i)$ is the prior probability that the target belongs to class $i$. 

\subsection{Data association of recognized features}
In any centralized or distributed tracking system, it is important to decide which of the received measurements are associated to the same target. As the reports may or may not contain the geo--location of the target, the data association algorithm should also be able to decide by considering only the features and probabilities that can be combined with each other. For instance, larger ships are associated with different types of emitter IDs and speed limits. This prior information must be available at any stage where association is necessary and processing must be applied to the received data to group them correctly with a common ID. The problem of association can also be addressed by the bearing and possibly range reports from ESM sensors.

\subsection{Classification and identification}
This step is the last processing block before the data is shown to the operator. If the data collected is in the form of probabilistic data, then the Bayesian framework is used for classification. However, if it is in the form of lists, for instance, belief functions are adopted to combine the data together. This can be done using Dempster's rule of combination.

%% file: systems.tex
\section{Hierarchical design of modern fusion systems} \label{new_idea}
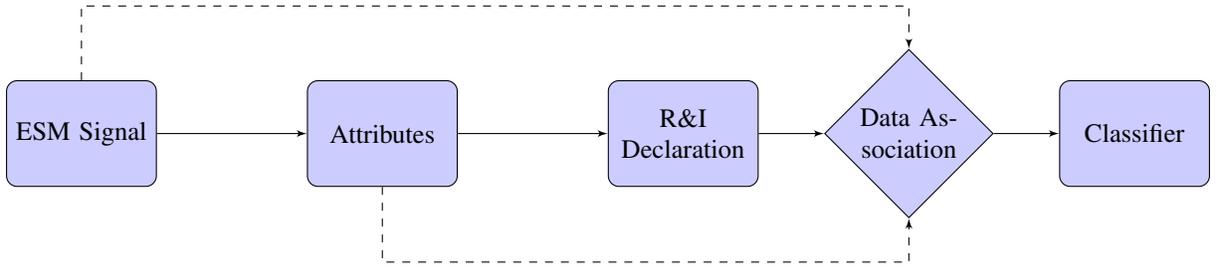
\begin{figure*}[!htbp]
\centering
\tikzstyle{decision} = [diamond, draw, fill=blue!20, 
    text width=4.5em, text badly centered, node distance=3cm, inner sep=0pt]
\tikzstyle{block} = [rectangle, draw, fill=blue!20, 
    text width=5em, text centered, rounded corners, minimum height=4em]
\tikzstyle{line} = [draw, -latex']
\tikzstyle{cloud} = [draw, ellipse,fill=red!20, node distance=3cm,
    minimum height=2em]
 \begin{tikzpicture}[node distance = 4cm, auto]
    \node [block] (init) {ESM Signal};
    \node [block, right of=init] (esm) {Attributes};
    \node [block, right of=esm] (att) {R\&I Declaration};
    \node [decision, right of=att] (decide) {Data Association};
    
    \node [block, right of=decide, node distance=3cm] (stop) {Classifier};
    \path [line] (init) -- (esm);
    \path [line] (esm) -- (att);
    \path [line] (att) -- (decide);
    \path [line] (decide) -- node { }(stop);
    \path [line,dashed] (esm.south) -| ([xshift=0cm,yshift=-1cm]esm.south) -| (decide);
    \path [line,dashed] (init.north) -| ([xshift=0cm,yshift=1.7cm]init.mid) -| (decide);
\end{tikzpicture}
\caption{Different levels of fusion for ESM/Tracker}
\label{fig_block2}
\end{figure*}

Because there exist different information types extracted at different levels of data processing, it is important to design a hierarchical fusion system that can work with information from different steps in Figure \ref{fig_block}. The revised architecture is shown in Figure \ref{fig_block2}.
According to Figure \ref{fig_block2}, other than the main flow of information (solid lines), there exist other possibilities to pass on the information at different stages. This is important when dealing with multiple ESM reports. In the next subsection, such a formulation is given for further development.

\subsection{Fusion of heterogeneous ESM information}
To be able to explain the fusion process, a simple example is suggested here. Assuming a scenario with two ESM sensors, the problem is to fuse the information if they are not processed to the same level and also vary in terms of the features they carry. Note that if all the sensors send a set that includes speed information or its related features, then one can easily fuse them as they can map, with some approximations, to each other. The problem arises when they cannot be mapped into each other and yet it is desired to fuse all the information before classification. Such situations can happen when, for example, one ESM sends a report about speed and the other one send information about the threat category of the ship. These ESM datasets do not carry the same information, and they are processed at different levels. For an ESM sensor, the report can be of any of the form
\begin{eqnarray}
\left\{ \mathbf{y}_{i},\mathbf{U}_{i}\right\} , & \left\{ \mathbf{a}_{j},\mathbf{A}_{j}\right\} , & \left\{ \mathbf{r}_{k},\rho_{k}\right\} \label{eq:ehsan_esm_sensors}
\end{eqnarray}
where different indexes show the possibility of different selected features that cannot be mapped to others directly. Assuming that there are $m_{\mathrm{ESM}}$ ESM sensors in the surveillance region, $m_{\mathrm{ESM}}^3$ possible combination can happen. If $m_{\mathrm{ESM}}=2$, then one possible combination is the set
\begin{equation}
\mathcal{S}=\left\{ \left\{ \mathbf{a}_{j},\mathbf{A}_{j}\right\} _{1},\left\{ \mathbf{r}_{k},\rho_{k}\right\} _{2}\right\} \label{eq:ehsan_ESM_2}
\end{equation}
where the indexes outside the brackets show the sensor number. After associating data into different groups for different objects or targets, the new classifier must be able to handle the heterogeneous flow of information. Defining a classifier function as
\begin{equation}
\Psi\left(\mathcal{S}_{1},\mathcal{S}_{2},\ldots,\mathcal{S}_{m_{\mathrm{ESM}}}\right)\label{eq:ehsan_class}
\end{equation}
the final output would be a new set of features, possibly similar to the R\&I declaration output to be sent to the operator.

One way to accommodate the fusion and classification of received data is to fill all the empty feature sets from collected data by different sensors if there exist non--overlapping data. In the case of overlapping, a simple fusion rule can be implemented as the data are of the same type and independent. After this processing, if there still exist some information gap to map all the data to the same feature set (R\&I declaration), a new classification function must be used to deal with it properly.

%% file: discussion.tex
\section{Simulation Results and Discussion} \label{discussion}
\begin{figure}[htbp]
    \centering
        \includegraphics[clip, trim=6cm 4cm 18cm 6cm,width=3in]{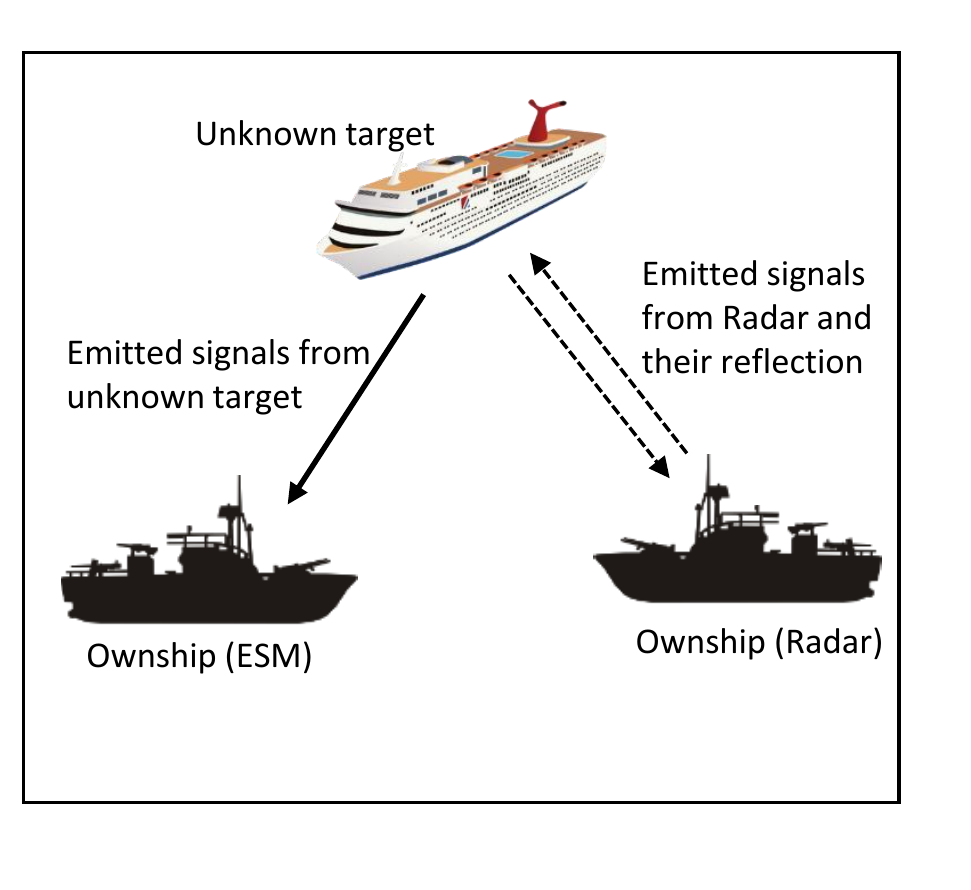}
    \caption{Maritime surveillance region}
    \label{fig_scenario}
\end{figure}

Now that the mathematical background and the system design of the sensor network have been defined, a simple example is discussed in this section to better understand the stpdf with R\&I of ESM and kinematic data. Here a distributed R\&I system is considered with two sensors, an ESM sensor and a radar. It is also assumed that a single ship is moving in the surveillance region as shown in Figure \ref{fig_scenario}. To illustrate the impact of  different processed data on the final classification output, it is assumed that the ESM report can be either the amplitude or the length. It is also assumed that the processed data from radar carries position and velocity in Cartesian coordinates.

The kinematics of the unknown target are sent to the fusion center as the kinematics vector $\mathbf{x}\left(k\right)$ and its associated covariance matrix $\mathbf{P}\left(k\right)$. The amplitude measured by ESM is sent as $\alpha\left(k\right)$ and the length as $L\left(k\right)$, respectively. In the simulations, it is assumed that the kinematics are processed in the Bayesian framework and the noises are white and Gaussian, hence a covariance matrix contains the full information about the uncertainty of the reported kinematics. The amplitude has Rayleigh distribution \cite{papoulis2002probability} and length has Gaussian distribution. The ESM reports, which are from different processing levels, are used for identification and classification separately and also together with the kinematics report to show the effect of adding ESM information for R\&I.

\begin{table}[htbp!]
\caption{Classes of the selected features}
\begin{centering}
\begin{tabular}{|c|c|c|c|}
\hline 
 & Speed ($v$) & $\alpha$ & $L$\tabularnewline
\hline 
\hline 
Class $1$ & $\mathcal{N}\left(5,3^{2}\right)$ & $\mathrm{Ray}\left(4\right)$ & $\mathcal{N}\left(15,2^{2}\right)$\tabularnewline
\hline 
Class $2$ & $\mathcal{N}\left(15,3^{2}\right)$ & $\mathrm{Ray}\left(2\right)$ & $\mathcal{N}\left(10,2^{2}\right)$\tabularnewline
\hline 
Class $3$ & $\mathcal{N}\left(30,3^{2}\right)$ & $\mathrm{Ray}\left(0.5\right)$ & $\mathcal{N}\left(5,2^{2}\right)$\tabularnewline
\hline 
\end{tabular}
\par\end{centering}
\label{tab_class_par}
\end{table}

The classes used in the simulations and the parameters assigned to each feature are shown in Table \ref{tab_class_par}. Note that each scenario only uses one of the ESM features to fuse with the kinematic (velocity) feature. The initial speed of the unknown target is $28{\mathrm{m}}/{\mathrm{s}}$ in Cartesian coordinates and the coarse is $\pi/4\:\mathrm{rad}$. The amplitude intensity is Rayleigh distributed with parameter $\sigma=0.5$ and the length of the ship is $5\mathrm{m}$. Also, the standard deviation of the uncertainty in the ship length's measurement is $5\mathrm{m}$. Taking in account all the information, it is expected that the data will be assigned to the third class.

\begin{figure}[htbp!]
    \centering
        \includegraphics[width=3.2in]{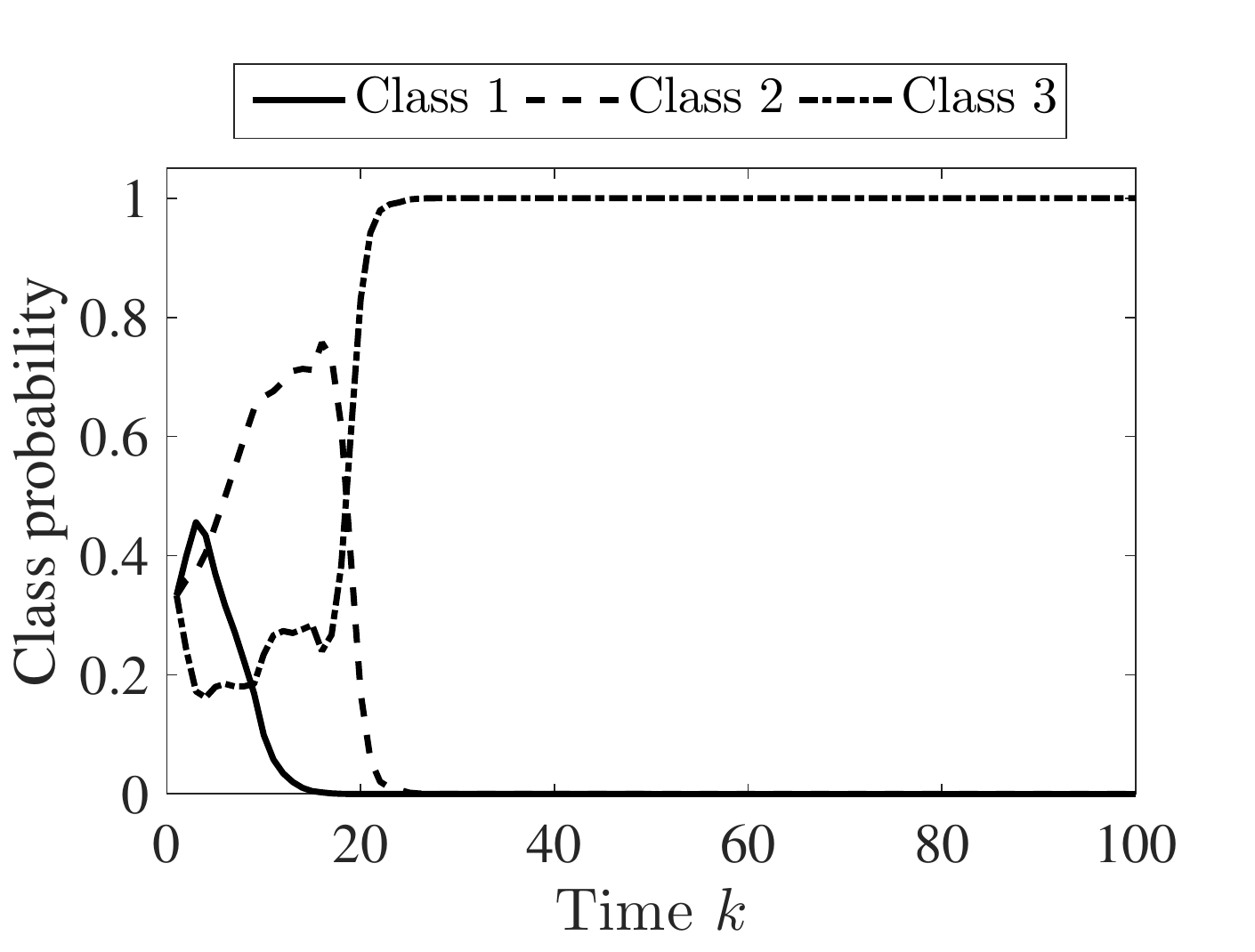}
    \caption{Kinematic feature class probabilities}
    \label{fig_kinematics}
\end{figure}

\begin{figure}[htbp!]
    \centering
        \includegraphics[width=3.2in]{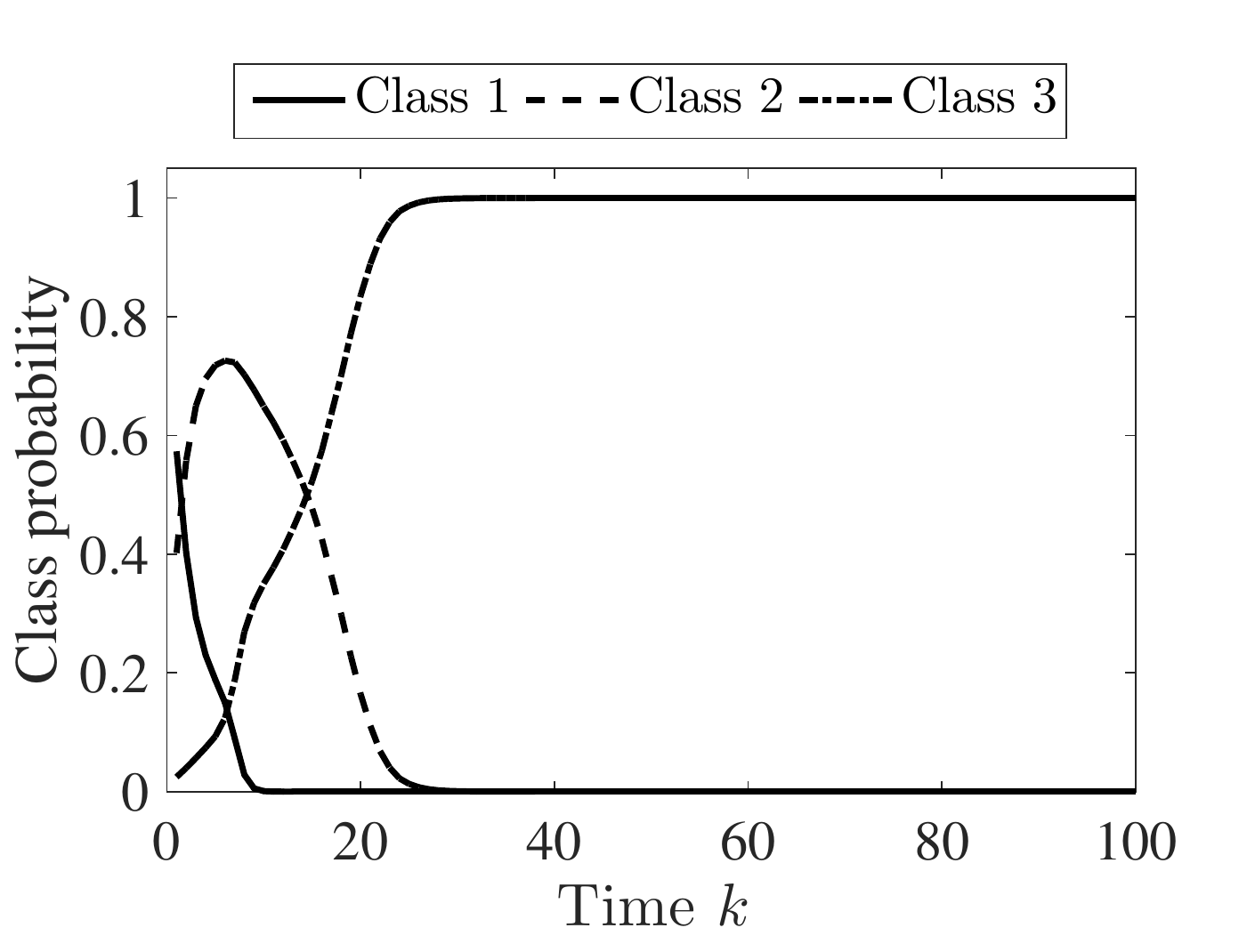}
    \caption{Ship amplitude features class probabilities}
    \label{fig_kinematics_a}
\end{figure}

\begin{figure}[htbp!]
    \centering
        \includegraphics[width=3.2in]{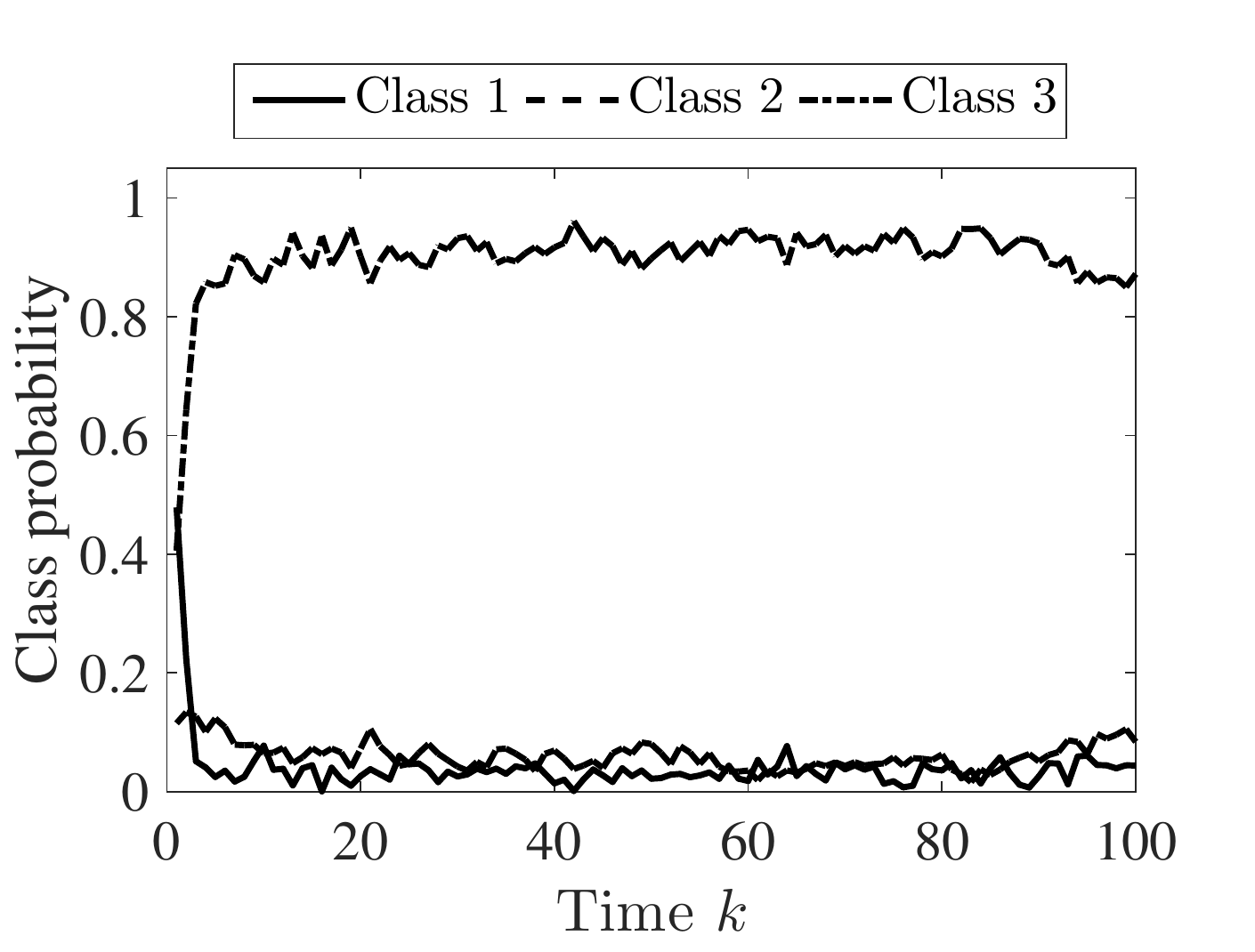}
    \caption{Ship length feature class probabilities}
    \label{fig_kinematics_L}
\end{figure}

\begin{figure}[htbp]
    \centering
        \includegraphics[width=3.2in]{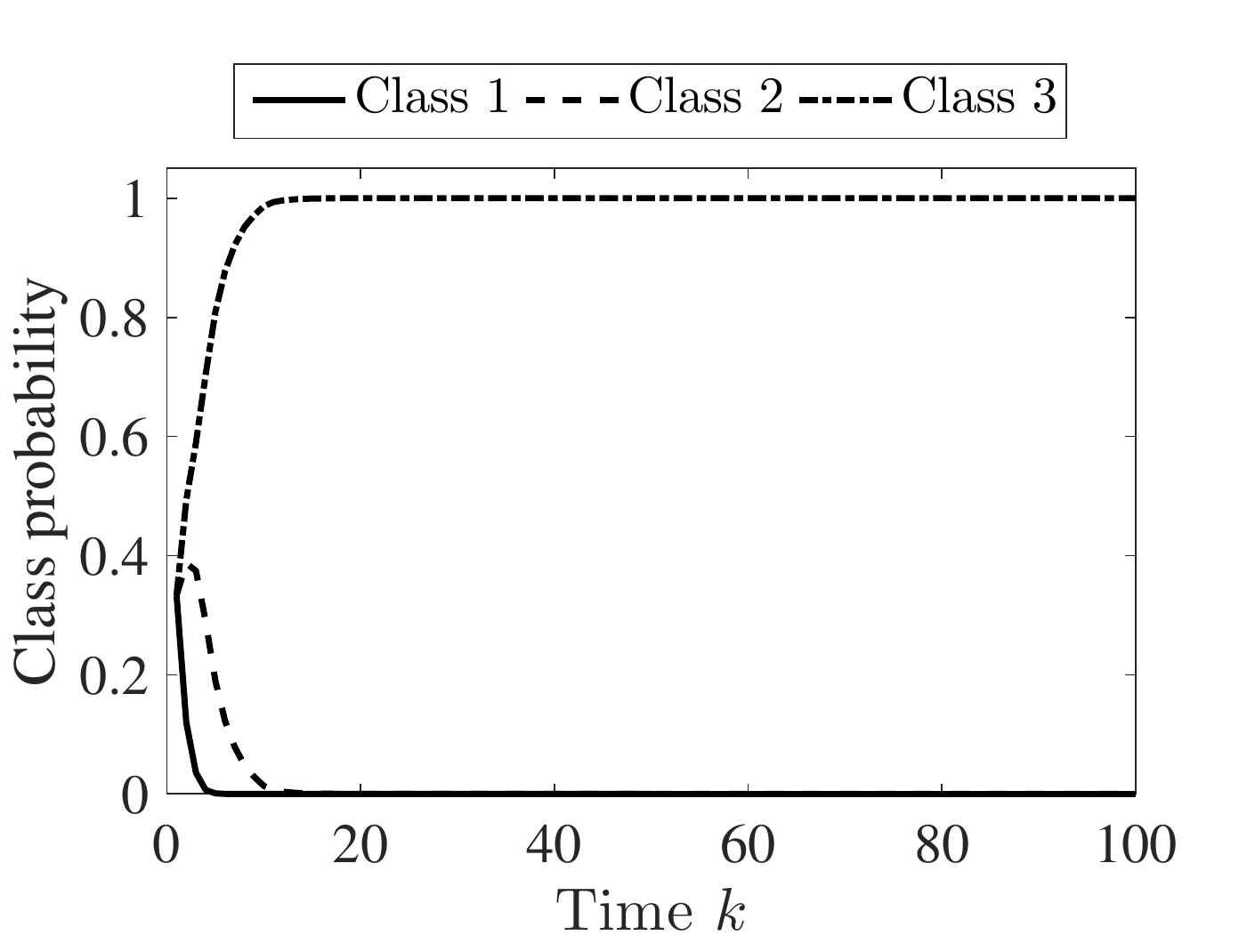}
    \caption{Kinematic and ship amplitude features class probabilities}
    \label{fig_kinematics_va}
\end{figure}

\begin{figure}[htbp]
    \centering
        \includegraphics[width=3.2in]{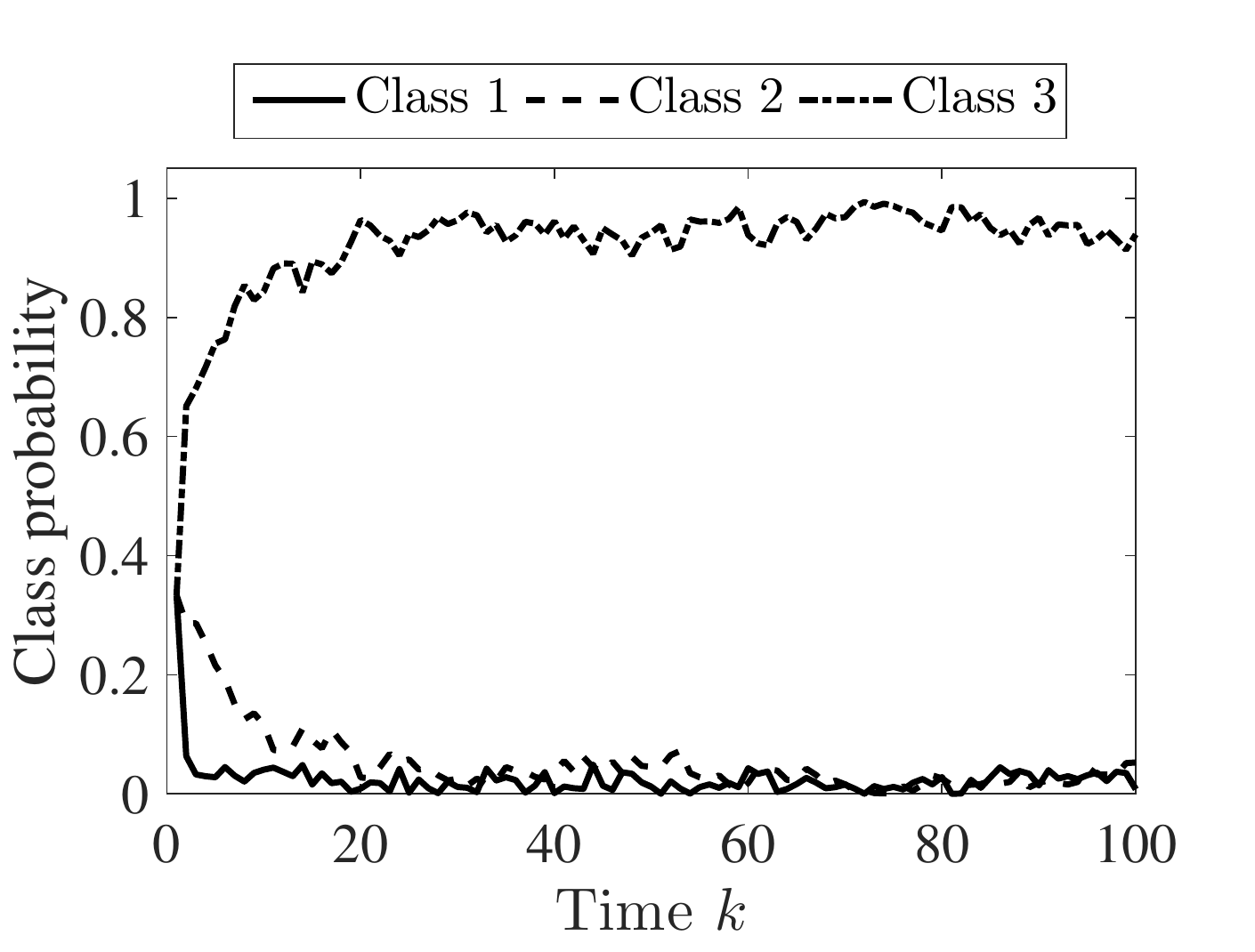}
    \caption{Kinematic and ship length features class probabilities}
    \label{fig_kinematics_vL}
\end{figure}

\begin{figure}[htbp]
    \centering
        \includegraphics[width=3.2in]{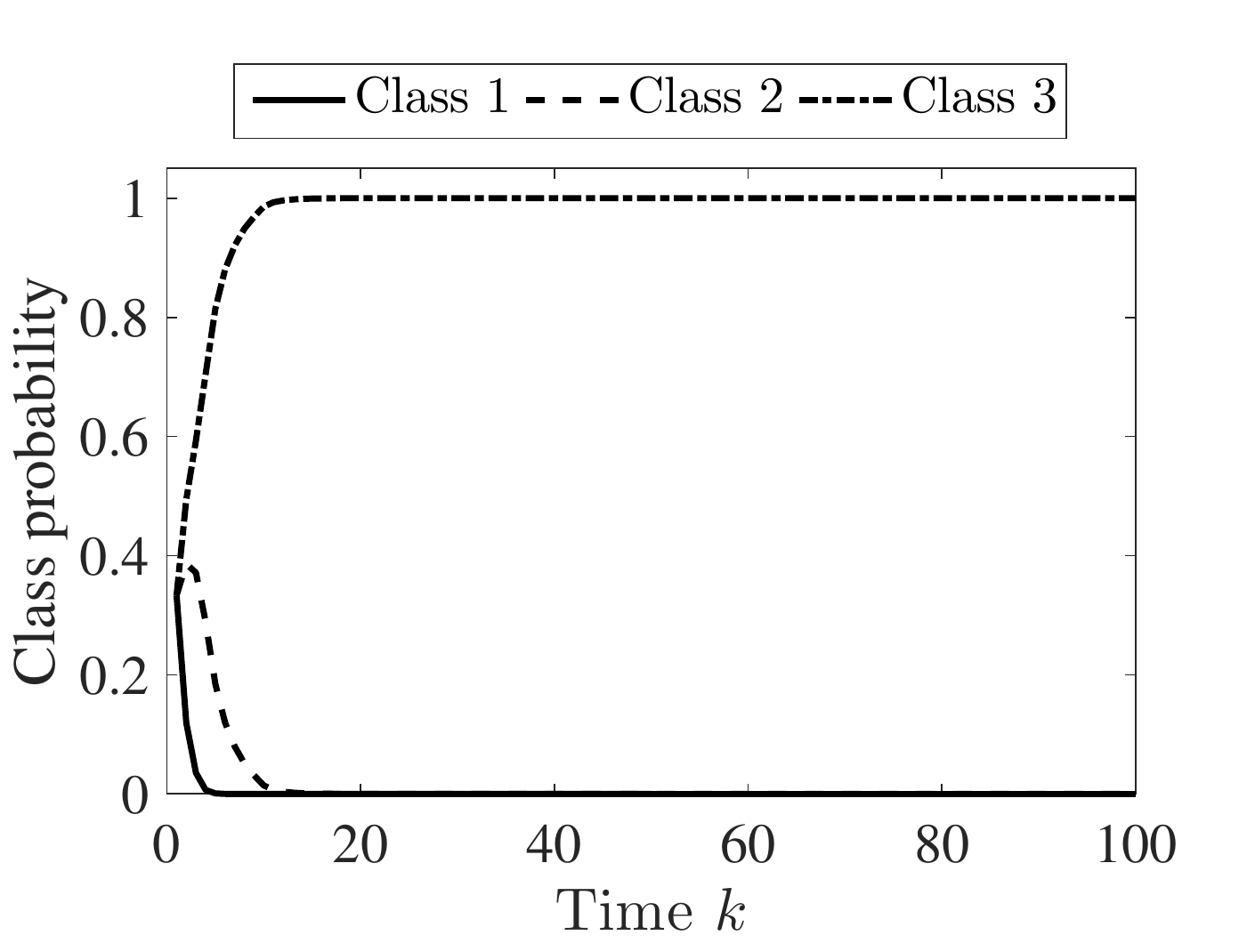}
    \caption{Kinematic, ship amplitude and length features class probabilities}
    \label{fig_kinematics_vaL}
\end{figure}

The classification is done over $100$ Monte Carlo runs for a scenario of $100$ time stpdf of reporting values from ESM and radar. Here, two important measures are considered to compare performance of each data fusion process. The first is the average of class probabilities over time (Figures \ref{fig_kinematics}--\ref{fig_kinematics_vaL}) and the second is the percentage of correctly reported class (Table \ref{tab_percent}).


Figures \ref{fig_kinematics}--\ref{fig_kinematics_L} show the class probabilities when only kinematic, length and amplitude features used separately for classification, respectively. In Figure \ref{fig_kinematics}, it can be seen that for the first few time stpdf, the class assignment is not stable. This shows that classifying based on only the kinematic of the unknown target can be misleading. Figure \ref{fig_kinematics_a} shows a more consistent result in terms of classification, however, the results from first twenty time stpdf are not reliable. Finally, Figure \ref{fig_kinematics_L} shows a more consistent result in terms of classification but with consistently lower probability assigned to the correct class. In order to classify the unknown target better, kinematic features can be coupled with either the data from the length of the ship (attribute feature), or the amplitude (signal level data from ESM). The results of these simulations are shown in Figures \ref{fig_kinematics_va} and \ref{fig_kinematics_vL}.
As shown in Figures \ref{fig_kinematics_va} and \ref{fig_kinematics_vL}, adding another feature to the classification can help the classifier distinguish between the unknown target features and other classes with higher accuracy. It is worth noting that amplitude and length information are from two different processing levels as stated in Figure \ref{fig_block2}. This shows that different ESM data, either attributes, signal level data or declared R\&I data can be coupled with kinematics features for classification. In order to demonstrate what is the best performance that can be achieved for the classification of the unknown target, all three features are combined together to classify the target and the results are shown in Figure \ref{fig_kinematics_vaL}. Here, the two ESM features that are contributing are processed differently. This shows that it is possible to gain more information and to classify the targets with more accurately even if data are not homogeneous.
It is important to know how many of the class reports were correctly in the simulations. This measure can be shown by counting the number of correct reported class in each Monte Carlo run and then take the average from them in each case. The results of such calculation are shown in Table \ref{tab_percent} for all the cases considered in Figures \ref{fig_kinematics}--\ref{fig_kinematics_vaL}. 
\begin{table}[htbp!]
\caption{Percentage of correct reported class (Class $3$)}
\begin{centering}
\begin{tabular}{|c|c|}
\hline 
Feature & Percent\tabularnewline
\hline 
\hline 
Velocity ($v$) & $82\%$\tabularnewline
\hline 
Amplitude ($\alpha$) & $87\%$\tabularnewline
\hline 
Length ($L$) & $90\%$\tabularnewline
\hline 
$v+\alpha$ & $98\%$\tabularnewline
\hline 
$v+L$ & $92\%$\tabularnewline
\hline 
$v+L+\alpha$ & $99\%$\tabularnewline
\hline 
\end{tabular}
\par\end{centering}
\label{tab_percent}
\end{table}

As shown in Table \ref{tab_percent}, there is a significant advantage in using length or amplitude features in conjunction with the kinematic data. As the numbers suggest, using a single feature does not yield highly accurate classification in all the cases. It is worth noting that in two of the cases, i.e., $(v+\alpha)$ and $(v+L+\alpha)$, the true class is assigned to the target in almost all the time stpdf.

The simulation shows that it is not necessary to consider only homogeneous ESM data, and different processing levels can be assumed for the system design. However, considering different processing levels should be done with some considerations for the type and uncertainty model of the data. One suggestion is to consider a combination of the Bayesian framework and belief functions to cover many of the situations that can occur in R\&I using ESM data. To do so, a more general list of all possible signal level, attributes and R\&I declared data must be available and suitable models should be assigned to them. Then, based on the suggested models, fusion and classification algorithms must be designed to address the problem in a consistent manner. This idea is in progress now.

%% file: conclusions.tex
\section{Conclusions} \label{conclusion}
In this paper fusion of ESM and kinematic data was discussed along with proposal for a general architecture for classification and identification of ESM data. In order to comply with the requirements of R\&I standards like STANAG $4162$, a hierarchical system design was proposed and further explored in the paper. The idea was to use different types of data from different processing levels and then combine them together to achieve better classification results. Simulation results showed that even adding one extra feature, from any level in the ESM processing chain, could improve the classification results solely based on kinematic.  

Unless the fusion of heterogeneous ESM and kinematic data is considered in modern system design, the performance of the feature extraction and classification of modules will be limited. To achieve better performance, multisensor solutions must be considered using the architectures proposed in this paper that utilize non--homogeneous data. Processing, R\&I and classification of non--homogeneous data from different sensors in realistic maritime solutions is in progress with consideration for the correlation among different levels of ESM data.